\documentclass{optica-article}
\journal{opticajournal} 
\articletype{Research Article}

\usepackage{lineno}
\usepackage{ulem}
\usepackage{xcolor}

\usepackage{siunitx}

\newcommand{\SiN}[0]{Si$_3$N$_4$}
\newcommand{\chitwo}[0]{$\chi^{(2)}$}

\begin{document}

\title{Broadband spectral mapping of photo-induced second-harmonic generation in silicon nitride microresonators}

\author{Ji Zhou,\authormark{1, *} Marco Clementi,\authormark{1, 2,*} Samantha Sbarra,\authormark{1} Ozan Yakar,\authormark{1} and Camille-Sophie Brès\authormark{1,$\dag$}}

\address{\authormark{1}Photonic Systems Laboratory (PHOSL)\char`,{} École Polytechnique Fédérale de Lausanne\char`,{} 1015 Lausanne\char`,{} Switzerland\\
\authormark{2}Present address: Dipartimento di Fisica "A. Volta", Università di Pavia, Via A. Bassi 6, 27100 Pavia, Italy\\
\authormark{*}These authors contributed equally to this work\\}

\email{\authormark{$\dag$}camille.bres@epfl.ch}

\begin{abstract*} By employing a pump-probe technique for enhanced spectral mapping of the dynamics in nonlinear frequency conversion, we demonstrate that photo-induced second-harmonic generation (SHG) in silicon nitride (\SiN) microresonators can persist when transitioning from the preferred doubly resonant condition--where the resonances of the optical harmonics are required to be matched--to a highly detuned state where the generated second harmonic is significantly shifted away from its corresponding resonance. This results in an unconventionally broad conversion bandwidth.  Other intriguing phenomena, such as detuning-dependent all-optical poling and nonlinear multi-mode interaction, are also presented for the first time with direct experimental evidence. 
Our findings provide new insights into the physics of photo-induced second-order (\chitwo) nonlinearity, highlighting its potential applications for nonlinear \chitwo\ photonics in integrated \SiN\ platform.
\end{abstract*}

\section{Introduction}

\noindent High-quality factor (Q) microresonators are attractive for efficient second-harmonic generation (SHG), owing to the ability to enhance the nonlinear light-matter interaction through tight optical confinement while enabling versatile dispersion engineering. 
Various material platforms have been extensively investigated for pushing the nonlinear conversion efficiency (CE), such as aluminum nitride \cite{Guo2016Secondharmonicgeneration, Bruch201817000}, lithium niobate \cite{Lu2019Periodicallypoledthin, Lu20201singlephoton}, silicon carbide \cite{Lukin2020}, or silica \cite{Zhang2019Symmetrybreakinginduced}. 
However, the high-Q nature, strict phase-matching conditions, and fabrication tolerance constrain the achievable SHG bandwidth, necessitating precise temperature control \cite{Surya2018Controlsecondharmonic}, the use of an auxiliary laser \cite{Hu2020Allopticalthermal}, dispersion compensation from coupling-induced frequency splitting \cite{Wang2020Strongnonlinearoptics}, or hybrid thermal- and electro-optic tuning \cite{briggs2024precise} to enable broadband nonlinear wavelength conversion or wavelength-accurate frequency doubling.

One remarkable exception is SHG in \SiN\ microresonators \cite{Nitiss2022Opticallyreconfigurablequasi, Hu2022Photoinducedcascaded, Li2023Highcoherencehybrid, Clementi2023chipscalesecond, sbarra2024uv}, where the observed 10-dB SHG bandwidth can significantly exceed the pump resonance linewidth. 
This intriguing phenomenon is counterintuitive. While the phase matching bandwidth, determined by the waveguide group velocity mismatch \cite{Jankowski2020Ultrabroadbandnonlinearoptics}, defines how many resonances can be involved in the SHG process, efficient generation is generally limited to the resonance linewidth, thereby restricting the process to a single resonance \cite{Guo2016Secondharmonicgeneration, Lu2020Efficientphotoinducedsecond}, unless delicate dispersion engineering on the waveguide group velocity mismatch is applied \cite{Hickstein2019Selforganizednonlinear}. Practically, single-resonance SHG is further constrained by thermal and Kerr effects, with only the pump remaining thermally locked to its nearest resonance, while the second harmonic (SH) is shifted away from its corresponding resonance \cite{Nitiss2022Opticallyreconfigurablequasi}. To the best of our knowledge, the extreme case of highly detuned SHG in high-Q \SiN\ microresonators has, thus far, only been predicted theoretically (SH detuning up to five linewidths) \cite{zhou2024sel}, with indirect experimental evidence that remains unexplained to date \cite{Nitiss2022Opticallyreconfigurablequasi, Hu2022Photoinducedcascaded, Li2023Highcoherencehybrid, Clementi2023chipscalesecond, sbarra2024uv}. 
This issue restricts the application of photo-induced second-order nonlinearity (\chitwo), particularly in combining third-order nonlinearity ($\chi^{(3)}$) based octave-spanning comb generation \cite{Li2017Stablyaccessingoctave, Pfeiffer2017Octavespanningdissipative, Briles2018InterlockingKerrmicroresonator, Weng2022Dualmodemicroresonators, Moille2023Kerrinduced, Zang2024Foundrymanufacturingoctave} and SHG-based comb self-referencing \cite{Okawachi2018Carrierenvelopeoffset, Hickstein2019Selforganizednonlinear, Nitiss2020Broadbandquasiphase} within the well-established \SiN\ integrated platform. These two nonlinear processes have been envisaged within a single \SiN\ microresonator \cite{Xue2017Solitontrappingcomb}, to facilitate compact on-chip all-in-one optical clocks. However, achieving SHG within the existence range of the soliton comb is challenging if only the doubly resonant condition is considered, due to the significant thermal/Kerr shift from the high pump power required for comb generation \cite{Guo2017universaldynamics}.

In this work, a highly detuned photo-induced SHG in \SiN\ microresonator is presented. By developing a novel pump-probe spectral mapping technique, we can track the detunings of optical harmonics during SHG over an unprecedented frequency range exceeding 10 GHz---more than tens of the SH resonance linewidth. We observe smooth transitions between the doubly resonant state and the highly detuned state. In the latter scenario, the generated SH exhibits substantial detunings, with no significant reduction in the generated SH power. We provide a qualitative explanation for this broadband reconfigurable SHG, based on previous theoretical understanding \cite{Dianov1995Photoinducedgenerationsecond, Yakar2022GeneralizedCoherentPhotogalvanic, zhou2024sel}.

\section{Theory of pump-probe technique}

\begin{figure*}[h]
    \centering
    \includegraphics[width=1\textwidth]{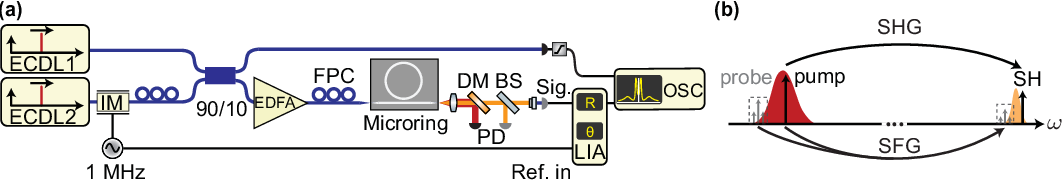}
    \caption{Characterization setup of the photo-induced second-harmonic using pump-probe technique.
    (a) Experimental setup for photo-induced SHG in \SiN\ microresonators and measuring the detunings of pump and SH resonances. ECDL: external-cavity diode laser; IM: intensity modulator; EDFA: erbium-doped fiber amplifier; FPC: fiber polarization controller; DM: dichroic mirror; PD: photodetector; BS: beam sampler; OSC: oscilloscope; LIA: lock-in amplifier.
    (b) Working principle of the pump-probe technique. The weak probe signals are up-converted via sum-frequency generation (SFG). The resulting beatings between the SFG signals illustrated within the dashed grey box at the SH resonance is picked up by the lock-in amplifier.
    }
    \label{fig1}
\end{figure*}

Fig. \ref{fig1}(a) presents the schematic of the experimental pump-probe setup used to track the detunings of the pump and its corresponding SH resonances. A weak probe laser (ECDL2) is modulated at a frequency of 1 MHz using an intensity modulator (IM) operating at the quadrature point, and then combined with the output of the pump laser (ECDL1). The combined signal is split into two branches. 
In the upper path, the electronic beating signal is filtered by a high-pass filter and serves as a marker for wavelength calibration between the two lasers. In the lower path, both the pump and probe signals are amplified by an erbium-doped fiber amplifier (EDFA) and then coupled to a \SiN\ microresonator. The \SiN\  microresonator features Q factors of approximately $0.7\times 10^6$ at telecommunication wavelengths. It is embedded in a silica cladding, with a waveguide cross-section of $1.7 \times 0.5$ \si{\micro\meter}$^2$ and ring radius of 158 \si{\micro\meter}. At the output of the \SiN\ microresonator, the collected light at SH band passes through a beam sampler (BS), is detected by a visible photodetector and sent to a fast lock-in amplifier, which is synchronized to the modulation frequency applied to the probe laser. 

The principle of the pump-probe technique is illustrated in Fig. \ref{fig1}(b). 
The wavelength of the modulated probe laser is scanned across that of the pump laser. 
As such, the sum-frequency generation (SFG) process takes place concurrently with SHG, similar to the degenerate SFG scheme shown in Ref. \cite{Wang2021EfficientFrequencyConversion}. The generated visible light at SH band is composed of both SFG and SHG signals. However, since the lock-in amplifier is referenced to the 1 MHz modulation frequency, only the beatings from the up-converted SFG signals are detected. 
Assuming an undepleted pump, we model the system with the following set of coupled mode equations: 
\begin{equation}
 \begin{aligned} \frac{\partial A}{\partial t}= & -\left(\frac{\kappa_{\mathrm{a}}}{2}+i \delta_{\mathrm{a}}\right) A +\sqrt{\kappa_{\mathrm{ex}, \mathrm{a}}} s_{\mathrm{in}}\left[1+\epsilon e^{-i \delta \omega t}\left(\frac{1}{2}+\frac{1}{4} e^{i \Omega t}+\frac{1}{4} e^{-i \Omega t}\right)\right] \\ \frac{\partial B}{\partial t}= & -\left(\frac{\kappa_{\mathrm{b}}}{2}+i \delta_{\mathrm{b}}\right) B+i g A^{2}\end{aligned} 
 \label{CME}
\end{equation}
where $A(B)$ represents the intracavity pump (SH) slowly varying field amplitude, normalized such that $ |A|^{2} $ and $ |B|^{2} $ correspond to the number of photons in the resonator. $\kappa_{\text{a(b)}}$ represents the total loss rate of the respective resonances. $\kappa_{\text{ex, a}}$ is the coupling strength of the pump resonance. $\delta_{\mathrm{a}}=\omega_{\mathrm{a}}-\omega_{\mathrm{pump}} $ and $ \delta_{\mathrm{b}}=\omega_{\mathrm{b}}-2 \omega_{\mathrm{pump}} $ are the effective detunings including thermal and Kerr shifts. $\omega_{\mathrm{a(b)}}$ and $\omega_{\text{pump}}$ are respectively the frequencies of pump (SH) resonance and pump laser. $s_{\text{in}}$ is the pump driving strength. $g$ is the nonlinear coupling strength related to the photo-induced \chitwo\ nonlinearity, which varies during system transitions \cite{Nitiss2022Opticallyreconfigurablequasi}. For the modulated probe laser, $\epsilon$ ($|\epsilon|\ll 1$) denotes the relative amplitude compared to the pump, and $\Omega$ is the modulation frequency. $\delta\omega=\omega_{\text {probe }}-\omega_{\text {pump}}$ is the tunable frequency difference between probe and pump lasers, which defines the detection range and can easily exceed 10 GHz.

For the system governed by Eq. (\ref{CME}), we consider the following ansatz: 
\begin{equation}
 \begin{array}{l}A=A_{0}+\epsilon e^{-i \delta \omega t}\left(A_{1}+A_{2} e^{i \Omega t}+A_{3} e^{-i \Omega t}\right) \\ B=B_{0}+\epsilon e^{-i \delta \omega t}\left(B_{1}+B_{2} e^{i \Omega t}+B_{3} e^{-i \Omega t}\right)+\mathcal{O}(\epsilon)\end{array} 
 \label{ansatz}
\end{equation}
where $A_i$ and $B_i$ ($i=0, 1, 2, 3$) are the complex amplitudes of the spectral components at each frequency. We focus on the pump SHG and pump-probe SFG (Fig. \ref{fig1}(b)), while neglecting higher-order terms of $\epsilon$ related to other nonlinear processes.

By substituting Eq. (\ref{ansatz}) into Eq. (\ref{CME}), we can derive the SH photodetector response $ i_{\mathrm{PD}} \approx \hbar 2 \omega_{\mathrm{pump}} \alpha r_{\mathrm{PD}} \kappa_{\mathrm{ex}, \mathrm{b}} |B|^{2}$, with $\hbar$ the reduced Planck constant, $\alpha$ the power collection efficiency of SH, $ r_{\mathrm{PD}} $ the responsivity of the SH photodetector, and $ \kappa_{\mathrm{ex}, \mathrm{b}} $ the coupling strength of SH resonance. Since the lock-in amplifier only detects the frequency components oscillating at $\Omega$, which correspond to the beatings between SFG signals $ 2 \operatorname{Re}\left\{\epsilon^{2}\left(B_{1} B_{3}^{*}+B_{1}^{*} B_{2}\right) e^{i \Omega t}\right\} $, with $\Omega\ll |\delta\omega|$ we obtain the lock-in amplitude response as $|H(\delta\omega, \Omega)| \sim i_{\mathrm{PD}}(\Omega)$ \cite{Nitiss2022Opticallyreconfigurablequasi}:
\begin{equation}
|H(\delta\omega, \Omega)| \approx \frac{P_{\mathrm{pump}}^{2}}{\hbar \omega_{\mathrm{pump}}} \cdot \frac{4 |\epsilon|^{2} \alpha r_{\mathrm{PD}} g^{2} \kappa_{\mathrm{ex}, \mathrm{a}}^{2} \kappa_{\mathrm{ex}, \mathrm{b}}}{\kappa_{\mathrm{a}}^{2} / 4+\delta_{\mathrm{a}}^{2}} \cdot \frac{1}{\left(\kappa_{\mathrm{b}}^{2} / 4+\left(\delta_{\mathrm{b}}-\delta \omega\right)^{2}\right)\left(\kappa_{\mathrm{a}}^{2} / 4+\left(\delta_{a}-\delta \omega\right)^{2}\right)}
\label{H_Omega}
\end{equation}
where $P_{\text {pump}}=\hbar \omega_{\text {pump}} |s_{\text {in }}|^{2}$ is the pump power in the bus waveguide. Note that here the derived response $|H(\delta\omega, \Omega)|$ is independent of the probe modulation frequency $\Omega$ under the above assumption, which is expected as we are probing the nonlinear response of the system without disturbing the underlying SHG dynamics. Eq. (\ref{H_Omega}) describes a two-peaked function, with the position of each (Lorentzian) peak corresponds to the detunings of the pump ($\delta_{\mathrm{a}}$) and SH ($\delta_{\mathrm{b}}$), and the linewidths of the peaks are associated with the respective loss rates. Therefore, a response map of the all-optical poling (AOP)-enabled SHG process can be obtained by continuously tuning the wavelengths of both the pump and probe lasers.
\begin{figure*}[t!]
    \centering
    \includegraphics[width=1\textwidth]{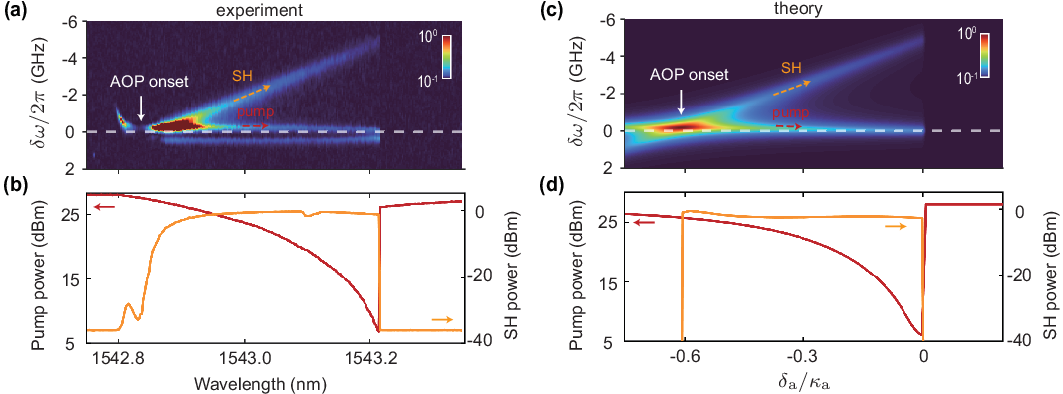}
    \caption{ Experimental (a-b) and theoretical (c-d) investigations of highly detuned second-harmonic generation involving TE$_{00}$ pump and TE$_{30}$ SH modes. (a) Measured lock-in response map when scanning the pump in and out of resonance. The response map is obtained by changing the pump laser wavelength in discrete steps, while the varying probe laser wavelength is adjusted accordingly. The triangular pump transmission, or thermal triangle, originates from pump-induced thermal and Kerr effects. 
    AOP, all-optical poling.
    (b) Transmitted pump power and the generated SH power in the bus waveguide as functions of pump wavelength.
    (c) Theoretically retrieved lock-in response map using Eq. (\ref{H_Omega}), with the detunings extracted from (a-b) for $\delta_{\mathrm{a,b}}<0$. Before the AOP onset ($\delta_{\mathrm{a}}/\kappa_{\mathrm{a}}<-0.6, \delta_{\mathrm{b}}>0$), the SH detuning is set to change linearly. The pre-poling dynamics before the AOP onset shown in (a) are excluded for simplicity. The red (yellow) arrows denote the pump (SH) power, respectively.
    (d) Simulated transmitted pump power and the generated SH power in the bus waveguide.}
    \label{fig2}
\end{figure*}

\section{Experimental results}

Fig. \ref{fig2} shows the typical complex dynamics of photo-induced SHG under pump wavelength tuning. The involved SH mode was confirmed as TE${_{30}}$ by two-photon imaging of the \chitwo\ grating \cite{Nitiss2022Opticallyreconfigurablequasi}. In repeated pump wavelength scans, when the pump is tuned into resonance, the \chitwo\ grating inscribed at 1543.22 nm from the previous scan supports SHG. As seen in Fig. \ref{fig2}(b), the generated SH power increases as the pump asymptotically approaches zero-detuning until a pump wavelength of 1542.81 nm, where the intracavity pump power becomes sufficient to modify the previously inscribed grating. Such `pre-poling' procedure effectively reduces the threshold pump power \cite{zhou2024sel} and has been applied in hybrid near-IR/visible low-noise sources via self-injection locking at low pump power \cite{Clementi2023chipscalesecond, Li2023Highcoherencehybrid}, as well as in spontaneous parametric down-conversion for entangled photon pair generation \cite{Li2025}. 

When the pump wavelength is further tuned beyond 1542.81 nm, the previous grating starts to be erased, resulting in a decrease in the SH power and the disappearance of the lock-in response (Fig. \ref{fig2}(a)). Beyond 1542.83 nm, a near doubly resonant condition sets in, triggering the inscription of a new grating (labeled as AOP onset) and rapid growth of SH power. Around 1543.10 nm, a dip is observed in the SH power trace (Fig. \ref{fig2}(b)), from the simultaneous green light emission we attribute it to nonlinear loss from the third-harmonic generation due to cascaded SHG and SFG processes \cite{Hu2022Photoinducedcascaded, sbarra2024uv, Li2025}. 

\begin{figure}[b!]
    \centering
    \includegraphics[width=1\textwidth]{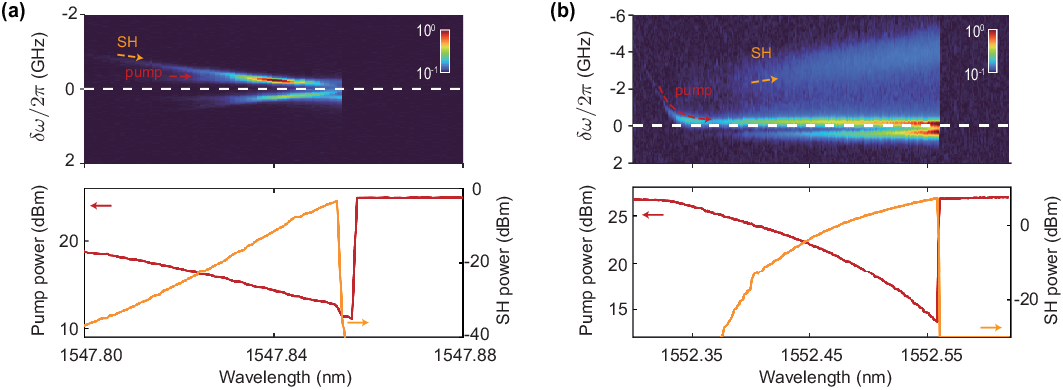}
    \caption{Lock-in response maps for SHG at different resonances in the same microresonator. (a) The case that the SH branch approaches the pump branch, and the SHG is prohibited when $\delta_{\text{b}}>0$. The red (yellow) arrows denote the pump (SH) power, respectively.  (b) The case where SHG involves the high-order SH mode (TE${_{40}}$), as indicated by the broad SH branch.}
    \label{fig3}
\end{figure}

\begin{figure*}[b!]
    \centering
    \includegraphics[width=1\textwidth]{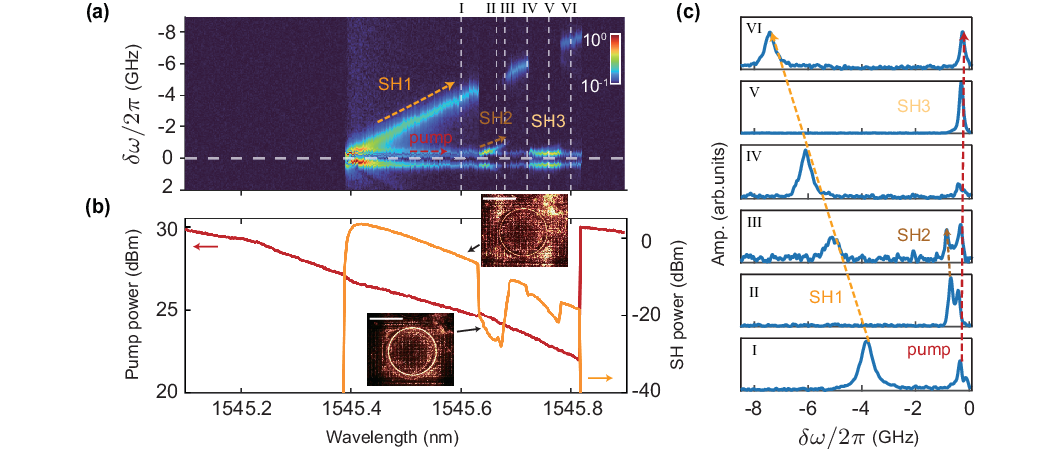}
    \caption{SHG involving multiple SH modes during a wavelength sweep in a pump resonance.
    (a) The SH1 branch is interrupted when the doubly-resonant conditions of the other two SH modes are successively satisfied. The SH3 branch is indistinct. 
    (b) The SH power trace exhibits discontinuities when SH mode competition or hopping occurs. Insets, the scattering patterns of the generated visible lights (scale bars, 200 $\si{\micro\meter}$). The red (yellow) arrows denote the pump (SH) power, respectively.
    (c) Lock-in responses at the linear scale, sliced from (a). The pump wavelengths (from $\rm{I}$ to $\rm{VI}$) are respectively 1545.60, 1545.67, 1545.68, 1545.72, 1545.76, and 1545.80 nm. In stage III, SHG occurs in both SH1 and SH2 modes.}
    \label{fig4}
\end{figure*}

Experimentally, we observe that the stimulated four-wave mixing process between pump and probe lasers is consistently present \cite{Wang2021EfficientFrequencyConversion}. As a result, the generated idler acts as a second probe, giving rise to additional corresponding lock-in responses for $\delta\omega>0$, as shown in Fig. \ref{fig2}(a), Fig. \ref{fig3} (a, b), and Fig. \ref{fig4}(a). However, we note that in all the experimental results presented here, only the pump branches are nearly mirrored. This is due to the fact that the idler is generally much weaker than the probe. Additionally, photo-induced resonant SHG requires both resonances to be blue-detuned ($\delta_{\text{a,b}}<0$) as a poling condition \cite{Carmon2004Dynamicalthermalbehavior, zhou2024sel}. Consequently, in the response maps for $\delta\omega>0$, the SFG between the weak idler and pump is too weak to replicate the SH branches from the probe. For simplicity, we have excluded it from our modeling in Eq. (\ref{CME}) and also the simulation in Fig. \ref{fig2}(b).  

The most notable characteristic of the reconfigurable SHG is its bandwidth of approximately 0.4 nm (50 GHz), limited by the thermal triangle length of the pump \cite{Carmon2004Dynamicalthermalbehavior}, with the potential to reach 0.6 nm (76 GHz) in the absence of additional perturbation from the probe laser \cite{Nitiss2022Opticallyreconfigurablequasi}. At the edge of the pump thermal triangle, the corresponding SH detuning is 4.8 GHz (Fig. \ref{fig2}(a)), yielding the CE is approximately $0.1\%/\mathrm{W}$, calculated from the pump power and generated SH power in the bus waveguide. Considering that the linewidths of pump and SH resonances are respectively 0.30 and 0.65 GHz estimated from Fig. \ref{fig2}(a), we conclude that such photo-induced SHG can still operate under a near-singly resonant condition where the generated SH is significantly detuned.

We provide a qualitative explanation for this broadband SHG phenomenon. Theoretically, photo-induced SHG in \SiN\ microresonators is constrained by threshold pump power \cite{Lu2020Efficientphotoinducedsecond, Lu2021ConsideringPhotoinducedSecond} and the detuning-dependent poling condition $\delta_{\text{a,b}}<0$ \cite{zhou2024sel}, while the quasi-phase-matching condition, corresponding to the period of the \chitwo\ grating, is automatically satisfied since the grating is inscribed by the light fields themselves \cite{Dianov1995Photoinducedgenerationsecond, Yakar2022GeneralizedCoherentPhotogalvanic}, and is given by $\Lambda=2\pi/(k_{\text{SH}}-2k_{\text{pump}})$, where $k_{\text{SH, pump}}$ are respectively the wavevectors of SH and pump. In the practical example shown in Fig. \ref{fig2}, the SHG is sustained as the system transitions from a doubly-resonant state to a highly detuned state, without resonant power enhancement for the SH field. Meanwhile, the out-coupled SH power remains nearly constant as a result of the compensation provided by the increasing intracavity pump power as its detuning decreases, as well as the optical dynamic reconfiguration of the \chitwo\ nonlinearity in both its amplitude \cite{Nitiss2020Broadbandquasiphase} and phase \cite{zhou2024sel}. This feature is well reproduced in the simulation of Fig. \ref{fig2}(d) using the model developed previously \cite{zhou2024sel}.

Unlike general SHG in other integrated platforms with intrinsic \chitwo\ nonlinearity, SHG in \SiN\ microresonators has been found to survive only with specific detunings of the involved optical harmonics. Specifically, a necessary condition for the existence of a stable nonlinear grating is that the pump/SH detunings satisfy $\delta_{\text{a,b}}<0$ \cite{zhou2024sel}. Here, we present the first experimental evidence of this poling condition, previously only theorized. This condition, already observed in Fig. \ref{fig2}(a), is further confirmed by the results shown in Fig. \ref{fig3}(a), which correspond to all-optical poling at another resonance involving a different SH mode from a low-order family, as inferred from its linewidth of around 0.1 GHz. In Fig. \ref{fig3}(a), the SH branch is observed to approach the pump branch when increasing the pump wavelength, in contrast to the scenario shown in Fig. \ref{fig2}(a). This opposite behavior can be explained as the frequency shift rate for this SH resonance being less than twice that of the corresponding pump resonance \cite{zhou2024sel, Zhang2019Symmetrybreakinginduced}. In this case, the SH power continuously grows due to the increased proximity to the doubly-resonant condition, achieving a CE of approximately $0.46\%/\mathrm{W}$ at the trailing edge of its trace. However, it suddenly drops once $\delta_{\text{b}}>0$. Therefore, SHG ceases before reaching the edge of the pump thermal triangle, despite the presence of high intracavity pump power.

Fig. \ref{fig3}(b) shows a similar case to Fig. \ref{fig2} but involving a lossy SH resonant mode with linewidth of approximately $\kappa_{\text{b}}/2\pi= 2.4$ GHz, which achieves a maximum CE of $2.2\%/\mathrm{W}$. Here, the exact doubly resonant condition is not seen at the onset of SHG due to the detuning-dependent threshold behavior of the AOP process \cite{zhou2024sel}, and the SH power continuously increases as the pump wavelength is increased. This dynamic behavior could not be observed in prior detuning measurements conducted with a vector network analyzer \cite{Nitiss2022Opticallyreconfigurablequasi}, where the detection range is limited by the 1 GHz bandwidth of the visible photodetector, restricting exploration to the doubly resonant states and their vicinity. The pump-probe technique presented here provides a significantly broader detection range, 
as well as improved sensitivity ($ |H(\delta\omega, \Omega)| \propto |\epsilon^2|$).

Finally, we can effectively investigate intriguing phenomena that occur exclusively in highly detuned states. For instance, it is likely that two neighboring SH resonances of different mode families can contribute to SHG simultaneously—--one enabling  a highly detuned SHG, while the other supporting  a doubly resonant SHG. Fig. \ref{fig4} presents the experimental results on SH mode competition, hopping and also coexistence. With higher pump power of around 30 dBm, the frequency offsets between SH resonances of different mode families can be compensated by thermal and Kerr shifts from the pump. We observe that SHG involving the SH1 mode (TE$_{20}$ \cite{Nitiss2022Opticallyreconfigurablequasi}) exhibits a continuous decrease in CE from $0.23\%/\mathrm{W}$, and this process can be interrupted by two other lower-order SH modes (labeled as SH2 and SH3), as evidenced by abrupt discontinuities in both the response map (Fig. \ref{fig4}(a, c)) and the SH power trace (Fig. \ref{fig4}(b)), alongside apparent changes in the scattering pattern of visible lights shown in Fig. \ref{fig4}(b) insets. Qualitatively, the dynamics of such SH mode switching and competition arises from the reconfigurability of AOP, which is activated, for individual modes, when the two aforementioned conditions are satisfied: i) $\delta_{\mathrm{b}} < 0$ and ii) the pump power is above the AOP threshold power. The threshold decreases with smaller detunings but does not require a strictly doubly resonant condition to trigger AOP. Consequently, AOP occurs simultaneously for multiple modes, naturally leading to SHG mode competition and hopping in such nonlinear system.

While theoretical modeling of SHG mode switching and competition remains an object of study \cite{zhou2024sel} and is beyond the scope of this work, these intriguing phenomena can be also understood qualitatively as inherent to the multimode nature of \SiN\ microresonators, which presents a fundamental duality. On the one hand, the variability of SH mode is often undesirable and can be suppressed through single- or few-mode waveguide geometry design that restricts the number of interacting transverse mode families. This approach can be further enhanced by employing refined control of the relative pump/SH mode detuning to enable ultrabroadband SHG at a desired fundamental mode \cite{Clementi2024UltrabroadbandResonantFrequency}. On the other hand, they can be harnessed to optically reconfigure the SH mode, thereby providing an additional degree of freedom to control the phase matching for other cascaded processes, such as  $\chi^{(3)}$ optical parametric oscillations involving different families of transverse modes \cite{sbarra2024uv}.

\section{Discussion and Conclusions}
Experimentally, to explore the limits of photo-induced SHG in the highly detuned regime, we purposely apply high pump power to extend the pump thermal triangle and induce significant shifts in the SH resonances. However, to optimize the SHG conversion efficiency, the chip temperature and pump power should be carefully adjusted to align the peak of SH power trace with the edge of pump thermal triangle, as efficient SHG favors the ideal doubly resonant condition. For further improvement of conversion efficiency, one could use smaller microresonators with high finesse \cite{Lu2020Efficientphotoinducedsecond}. In addition, coupled microresonators can offer more flexibility in resonance detunings and loss control compared to single microresonator, where the shifts of SH and pump resonances are coupled \cite{Clementi2024UltrabroadbandResonantFrequency, Zhao2021Theory2microresonator}. These approaches are beyond the scope of this work.

In conclusion, we demonstrated highly detuned SHG in a high-Q \SiN\ microresonator using a novel pump-probe technique, which facilitates the investigation of the broadband dynamics of resonant photoinduced SHG enabled by the AOP process. We unambiguously verified the all-optical poling condition $\delta_{\text{a,b}}<0$, under which the optical harmonics inscribe the \chitwo\ nonlinear grating at varying pump wavelengths in a reconfigurable fashion. As the system transitions with increasing detuning of generated SH, this gives rise to broadband SHG. 

We emphasize that the broadband SHG occurs without engineering the group-velocity matching of the resonator waveguide, which could be further optimized to achieve simultaneous broadband SHG at multiple resonances or even comb up-conversion \cite{Guo2018efficient, Hickstein2019Selforganizednonlinear, Clementi2024UltrabroadbandResonantFrequency, mehrabad2025multi}. 
Such optical reconfigurable SHG significantly relaxes the dispersion engineering requirements for SHG, showing compatibility with other nonlinear processes such as $\chi^{(3)}$ optical parametric oscillation \cite{sbarra2024uv}, coupled $\chi^{(2)}-\chi^{(3)}$ soliton combs generation  \cite{Xue2017Secondharmonicassisted,Hu2022Photoinducedcascaded,Talenti2025} even with relatively low photo-induced \chitwo\ nonlinearity \cite{Lu2020Efficientphotoinducedsecond, Nitiss2022Opticallyreconfigurablequasi}, and octave-spanning soliton comb generation \cite{Moille2023Kerrinduced}. 
In the latter case, the strong soliton pump inevitably introduces a thermal crosstalk and deviates the preferred doubly resonant condition for the SHG of a comb line at the spectral wing.  Our study demonstrates that SHG for $f-2f$ self-referencing remains feasible even in the presence of the thermal crosstalk from the strong soliton pump. Moreover, wavelength-precise SHG targeting specific atomic transitions is also achievable, significantly enhancing its applicability to optical clocks. Together, these capabilities open a path toward monolithic integration of octave-spanning comb generation and its carrier–envelope offset detection within a single \SiN\ microresonator, advancing the development of compact optical clocks \cite{Spencer2018, Xue2017Solitontrappingcomb, Moille2023Kerrinduced}.

\begin{backmatter}
\bmsection{Funding}  ERC grant PISSARRO (ERC-2017-CoG 771647); Swiss National Science Foundation (SNSF grant MINT 214889).

\bmsection{Acknowledgment} We thank Anton Stroganov from Ligentec SA for providing the \SiN\ chip. 

\bmsection{Disclosures} The authors declare no conflicts of interest.
\end{backmatter}

\bibliography{ref}

\end{document}